\documentclass[a4paper,reprint,nofootinbib,amssymb,amsmath,aip,groupedaddress]{revtex4-1}
\setcitestyle{numbers,square}
\usepackage{bm}
\usepackage{bbm}
\usepackage{braket}
\usepackage[colorlinks=true,linkcolor=blue]{hyperref}
\expandafter\ifx\csname package@font\endcsname\relax\else
 \expandafter\expandafter
 \expandafter\usepackage
 \expandafter\expandafter
 \expandafter{\csname package@font\endcsname}
\fi
\usepackage{amsmath}
\usepackage{hhline}

\usepackage{tikz}
\usepackage{slashed}

\definecolor{darkgreen}{HTML}{109930}

\newcommand{\mat}[1]{\left(\begin{matrix}#1\end{matrix}\right)}
\newcommand{\D}[1]{\ensuremath{\mathrm{D}_{#1}}}
\newcommand{\I}{\ensuremath{\mathrm{i}}}

\begin{document}

\title{Simultaneous Block Diagonalization of Matrices of Finite Order}%

\author{Ingolf Bischer}%
\email{bischer@mpi-hd.mpg.de}
\author{Christian D\"oring}%
\email{cdoering@mpi-hd.mpg.de}
\author{Andreas Trautner}%
\email{trautner@mpi-hd.mpg.de}
\affiliation{Max-Planck-Institut f\"ur Kernphysik, Saupfercheckweg 1, 69117 Heidelberg, Germany}%

\begin{abstract}
\textbf{Abstract.}
It is well known that a set of non-defect matrices can be simultaneously diagonalized if and only if the matrices commute. 
In the case of non-commuting matrices, the best that can be achieved is simultaneous block diagonalization. 
Here we give an efficient algorithm to explicitly compute a transfer matrix which realizes the simultaneous block diagonalization of unitary matrices 
whose decomposition in irreducible blocks (common invariant subspaces) is known from elsewhere. 
Our main motivation lies in particle physics, where the resulting transfer matrix must be known explicitly in order to unequivocally determine the action of 
outer automorphisms such as parity, charge conjugation, or time reversal on the particle spectrum.
\end{abstract}

\maketitle

\section{Introduction}\label{sec:intro}
A standard problem in group theory is the decomposition of matrix representations 
into their irreducible invariant subspaces (irreps).
Given an explicit matrix representation of a group,
the well-known character analysis, see e.g.\ \cite{Ramond:2010zz} for finite groups,
serves to determine the number and minimal size of blocks
that can be achieved in a simultaneous block diagonalization of all representation matrices.
However, it is in general a very different problem to perform
such a simultaneous block diagonalization explicitly.
In this note we introduce an algorithm that solves this problem.
Given a set of original input matrices in an arbitrary basis,
as well as their decomposition into irreducible blocks (i.e.~the number and dimension of the blocks),
the algorithm gives the transfer (basis-transformation) matrix
that rotates all matrices simultaneously to their block diagonal form. 

\medskip 

The algorithm presented here applies to any finite set of equal-dimension unitary matrices
for which the decomposition into minimal common invariant subspaces is explicitly known.
The problem of finding such common invariant subspaces and their explicit representations
seems to be generally solved only if the matrices obey a group structure,
and it is particularly tractable if the corresponding group is finite.
Nontheless, a method to find the number of $d$-dimensional common invariant subspaces of a finite set of 
square matrices for arbitrary $d$ was presented in~\cite{Arapura:2004}.
The \textit{existence} of a similarity transformation 
that performs the simultaneous block diagonalization of matrices with blocks of
dimension one or two has been addressed in \cite{Watters:1974} and \cite{Shapiro:1979},
while here we seek for an explicit derivation of this transformation,
also for blocks of arbitrary dimensions.
Our algorithm partly benefits from ideas set forth in \cite{Shemesh:1984}.

\medskip 

Applications of our algorithm and the resulting transfer matrix are manifold.
First and foremost, our algorithm can be used to explicitly find a simultaneously block diagonal basis for a finite group.
This problem is, for example, commonly encountered in the breaking of continuous groups to their finite subgroups~\cite{Adulpravitchai:2009kd, Luhn:2011ip, Merle:2011vy, Fallbacher:2015pga}.
But applications reach far beyond that. For instance, it may occur that there exist matrix operators living in the 
same space as the group matrices, but which are themselves not part of the group. 
If the action of the group is diagonalized it is often very important to know
how these operators transform, and this requires explicit knowledge of the basis transformation matrix.
In the continuous$\rightarrow$discrete example above, this would be operators that live in the coset of the continuous group with respect to the finite subgroup.
These operators correspond to non-linearly realized symmetries, implying that knowing their action on
the physical fields, which can be derived by our algorithm, is instrumental in constructing the low-energy effective field theory~\cite{Coleman:1969sm,Callan:1969sn},
see also~\cite{Das:2020arz} for a recent example.

Other examples for operators that act on the same space as the group, but which are themselves not part of the group,
are representations of outer automorphisms. 
For instance, knowing the transfer matrix is strictly required in order to compute the action of outer 
automorphisms on non-product representations\footnote{%
While our algorithm does apply to product representations, 
for them the problem of explicitly decomposing a representation into irreducible blocks
is solved, in general, by knowledge of the Clebsch-Gordan coefficients.} 
such as, for example, the regular representation of a finite group. 
This will be investigated in more detail by the present authors in a forthcoming
publication \cite{Bischer:2020xxx}. 
This is of particular physical interest, because parity (P), charge conjugation (C), and time reversal (T) 
transformations are all known to correspond to outer automorphisms \cite{Buchbinder:2000cq, Grimus:1995zi, Trautner:2016ezn},
which is specifically true for finite groups~\cite{Holthausen:2012dk}.
For these, physically very interesting situations can arise where CP is explicitly violated 
by the Clebsch-Gordan coefficients of a finite group~\cite{Chen:2009gf,Chen:2014tpa} or CP transformations
may be forced to be of order larger than two for specific groups~\cite{Chen:2014tpa} (see also~\cite{Chen:2019iup} for a brief introduction, 
and~\cite{Ivanov:2015mwl} for a specific phenomenological model). While standard bases for such general CP transformations~\cite{Lee:1966ik} exist, 
see~\cite{Ecker:1987qp} and also \cite[Ch.2,App.C]{Weinberg:1995mt}, such bases may typically not be attained while keeping 
the linear symmetry (block) diagonal. Our algorithm, here can be used to derive the 
action of a general CP transformation on the physical spectrum, 
which is obtained after diagonalizing the action of the linearly realized symmetry group.
Finally, the above examples also appear in combination: Breaking a continuous group to a finite subgroup
while tracking the effect of the physical CP transformation on the physical spectrum 
requires knowledge of the transfer matrix that diagonalizes the action of the linearly realized group~\cite{Ratz:2016scn}.

In many of the above examples, the explicit block diagonalization has been done manually  
which is not a real challenge for a small number of generators or small dimensional representations.
However, the problem becomes more complicated and quickly grows out of hand 
for more complicated situations with larger groups and/or larger representations.
The benefit of our algorithm is that it seamlessly extends to such situations.

\medskip

We will now present the algorithm in form of a proposal and subsequently prove it.
In App.~\ref{sec:D_8} we give an example based on the regular representation of the group~\D8.

 \section{Simultaneous matrix transformation}\label{sec:SimTrafo}
Consider a collection of $N$, $D$-dimensional unitary matrices $G_g\in \mathrm{U}(D)$, where $g=1,\dots,N$.
We assume that there exists a unitary (hence, invertible) $S\in  \mathrm{U}(D)$ such that
\begin{equation}\label{eq:1}
   S^{-1}\, G_g\, S = \mathcal{B}_g \qquad \forall g\in\{1,\dots,N\},
\end{equation}
where $\mathcal{B}_g$ is a collection of unitary block diagonal matrices. 
Further, we assume that the $\mathcal{B}_g$'s here are composed of 
blocks of \textit{minimal} size,
i.e.\ $S$ realizes a decomposition of linear 
transformations $G_g$ into their minimal common invariant subspaces.
In this work we give a fast constructive algorithm to explicitly obtain a matrix $S$ satisfying the above requirements.

Let us first establish a standard form of the matrices $\mathcal{B}_g$. Each $\mathcal{B}_g$ 
can be written as a direct sum of blocks $B$,
\begin{equation}\label{eq:sumofblocks}
    \mathcal{B}_g=B_g^1\oplus B_g^2\dots \quad \forall g.
\end{equation}
In general, it may occur that the $k$-th and $l$-th block in $\mathcal{B}_g$ are identical.
If such a degeneracy extends over all $g$ (for some fixed indices $k$ and $l$), that is, if
\begin{equation}\label{eq:degenBlocks}
B_g^k=B_g^l \qquad \forall g,
\end{equation}
we speak of degenerate blocks. 
Note that we may use \eqref{eq:degenBlocks} 
without loss of generality even if the blocks are only 
identical up to a global (i.e.\ $g$-independent) similarity transformation,
since such a transformation can always be absorbed in $S$.
We introduce \mbox{$b=1,...,n$} which runs over the blocks, counting 
degenerate blocks only once, and the numbers $q_b$ and $d_b$ 
for the multiplicity (i.e.\ the degeneracy) and dimensionality of a given block $b$, respectively.
Hence, by definition
\begin{equation}
 \sum_{b=1}^{n} q_b\,d_b=D.
\end{equation}
In our standard form, we order the direct sum \eqref{eq:sumofblocks} such that degenerate blocks appear in direct succession, i.e.\
\begin{equation}\label{eq:block-decomposition}
    \begin{split}
    \mathcal{B}_g&=\bigoplus_{b=1}^{n} \bigoplus_{q=1}^{q_b} B_g^{b}\equiv \bigoplus_{b=1}^{n} (B_g^{b})^{\oplus q_b} \\
     &= \underbrace{B_g^1\oplus \dots \oplus B_g^1}_{q_1\text{ times}} \oplus \dots\oplus \underbrace{ B_g^{n}\oplus\dots\oplus B_g^{n}}_{q_{n}\text{ times}}\qquad\forall g\,.
    \end{split}
\end{equation}

We now state the construction of $S$ in the form of a proposition and subsequently prove it.

\medskip 

\textit{Proposition 1.} Define
\begin{equation}\label{eq:BlockSolution}
 \mathcal{M}_g^b:=\left[(\mathbbm{1}_{d_b}\otimes G_{g})-(B_g^{b,\mathrm{T}}\otimes \mathbbm{1}_{D})\right],
\end{equation}
and
\begin{equation}\label{eq:defM}
\mathcal{M}^b := \mat{\mathcal{M}^b_1\\ \vdots \\ \mathcal{M}^b_N }\,.
\end{equation}
The kernel $\ker\mathcal{M}^b$ is $q_b$-dimensional, i.e.\ it can be spanned by $q_b$ 
orthogonal \mbox{$(D\cdot d_b)$-dimensional} vectors $w^{b}_{q=1,\dots,q_b}\in \mathbbm{C}^{(D\cdot d_b)}$.
A solution to \eqref{eq:1} then is given by\footnote{%
$\mathrm{vec}_D^{-1}$ converts a $D\cdot k$ dimensional vector into a $D\times k$ matrix by taking 
the first $D$ components as the first column, the second $D$ components as the second column and so forth.
}
\begin{equation}\label{eq:solS}
S = \mathrm{vec}_D^{-1}
\mat{
w^1_1 \\ 
\vdots \\
w^{1}_{q_1} \\[2pt]
w^{2}_1 \\
\vdots \\
w^{2}_{q_2} \\
\vdots \\ 
w^n_1 \\
\vdots \\
w^b_{q_n}}.
\end{equation}
$S$ is invertible with pairwise orthogonal columns, such that
we can always normalize them in order to promote $S$ to a unitary matrix.

\medskip

To prove our proposition, let us first reformulate Eq.~\eqref{eq:1}. 
We use the vectorization operation, which transforms an $n\times m$ matrix $A$ into an $nm\times 1$ vector $\mathrm{vec}(A)$ by stacking the columns of the matrix $A$ on top of each other. 
Given two matrices $X\in\mathbbm{C}^{a\times b}$ and $Y\in\mathbbm{C}^{b\times c}$ the vectorization of the product of the matrices fulfills the identities
\begin{align}\label{eq:VecId}
    \mathrm{vec}(XY)=(\mathbbm{1}_c\otimes X)\mathrm{vec}(Y)=(Y^{\mathrm{T}}\otimes \mathbbm{1}_a)\mathrm{vec}(X)\,.
\end{align}
Using these, we reformulate Eq.~\eqref{eq:1} as
\begin{equation}\label{eq:2}
 (\mathbbm{1}_D\otimes G_g)\mathrm{vec}(S)=(\mathcal{B}_g^{\mathrm{T}}\otimes \mathbbm{1}_D)\mathrm{vec}(S) \qquad \forall g.
\end{equation}
Note that Eq.~\eqref{eq:2} holds even if $S$ is not invertible.
The required invertibility of $S$ is kept in mind as additional information.
We then use \eqref{eq:block-decomposition} to decompose the matrix on the r.h.s.\ of Eq.~(\ref{eq:2}) as
\begin{equation}
 \left[\mathcal{B}^{\mathrm{T}}_g\otimes \mathbbm{1}_D\right]=
 \left[\bigoplus_{b=1}^{n}\left(B_g^{b,\mathrm{T}}\otimes \mathbbm{1}_D\right)^{\oplus q_b}\right]\;.
 \end{equation}
With regard to this, it makes sense to also write the matrix on the l.h.s.\ of Eq.~(\ref{eq:2}) as a direct sum,
\begin{equation}
\left(\mathbbm{1}_D\otimes G_g\right)= \left[\bigoplus_{b=1}^{n}\left(\mathbbm{1}_{d_b}\otimes G_g\right)^{\oplus q_b}\right]\;. 
\end{equation}
In this way, Eq.~(\ref{eq:2}) decomposes into blocks, with the smallest commensurable blocks
of both sides being of size $(d_b\cdot D)\times(d_b\cdot D)$, and those blocks appear with a multiplicity $q_b$.
Using this decomposition, we can rewrite \eqref{eq:2} as 
\begin{equation}\label{eq:MS}
\left[\bigoplus_{b=1}^{n}(\mathcal{M}_g^b)^{\oplus q_b}\right] \mathrm{vec}(S) = \vec0_{D^2}\qquad\forall g\;,
\end{equation}
with $\mathcal{M}_g^b$ as defined in \eqref{eq:BlockSolution}, and $\vec0_{D^2}$ being the \mbox{$D^2$-dimensional} null vector.
Due to the $q_b$ fold degeneracy in \eqref{eq:MS}
one actually just has to solve the equations 
\begin{equation}\label{eq:Mker}
 \mathcal{M}_g^b\,w^b = \vec 0_{(D\cdot d_b)}\qquad\forall g\;,\qquad \text{(no sum)}
\end{equation}
for a $(D\cdot d_b)$-dimensional vector $w^b$. 
That is, for each block~$b$ one has to find the intersection of the kernels of $\mathcal{M}_{g=1,\dots,N}^b$, 
i.e.\  the kernel of $\mathcal{M}^b$ as defined in \eqref{eq:defM}.
This kernel is $q_b$-dimensional,
because it is in a one-to-one correspondence with the according invariant subspaces of $G_{1,\dots,N}$, 
of which we have \textit{assumed} there exist $q_b$ copies. 
A more detailed proof of $\dim\ker\mathcal{M}^b=q_b$ is given in App.~\ref{sec:dimKernM}.
Therefore, for each block $b$ there are exactly $q_b$ non-trivial linearly independent solutions to \eqref{eq:Mker}. 
We choose a basis for these solutions, spanned by $q_b$ orthogonal vectors $w^b_{q=1,\dots,q_b}$.
Put together as in \eqref{eq:solS}, these form a non-trivial solution of \eqref{eq:2} and, for invertible $S$, 
also a solution of \eqref{eq:1}.

\medskip

We now discuss the conditions that $S$ obtained by our construction is invertible,
which turns out to be always the case given our assumptions.
The requirement that $S$ is invertible 
puts a stronger condition on the solution 
of \eqref{eq:Mker} than just the existence of $q_b$ linearly independent solutions. 
In order to formulate this, let us partition each $(D\cdot d_b)$-dimensional vector $w^b_q$ into $d_b$, $D$-dimensional vectors as
\begin{equation}\label{eq:winv}
w^b_q = \mat{v^b_{q,1} \\ \vdots \\ v^b_{q,d_b}}.
\end{equation}
Invertibility of $S$ now requires not only the $w^b_q$'s to be linearly independent,
but in fact, it requires that \textit{all} of the $v^{b}_{q,\beta}\in\mathbbm{C}^D$ ($b=1,\dots,n$, $q=1,\dots,q_b$, $\beta=1,\dots,d_b$) must be linearly 
independent, and, in particular, none of them can be zero. 
Clearly, if an invertible $S$ exists, we \textit{must} be able to find such a solution.

First, note that all non-trivial solutions to \eqref{eq:Mker} have the feature 
that the according $v^{b}_{q,\beta=1,\dots,d_b}$ are orthogonal. 
To see this, define a $(D\times d_b)$-dimensional matrix \mbox{$W:=\mathrm{vec}_D^{-1}(w^b_q)$}
and rewrite \eqref{eq:Mker}, using \eqref{eq:VecId} in reverse, as 
\begin{equation}\label{eq:GUUB}
 G_g\,W = W\,B_g^b\qquad\forall g\;.
\end{equation}
From invertibility of $G_g$ and $B_g^b$, and 
irreducibility of $B_g^b$ we find that $\ker W=0$ (another solution would be $W$ being the zero matrix, 
which is excluded by our desire to discuss a non-trivial solution of \eqref{eq:Mker}).
Consequently, $W$ is \mbox{left-invertible}, implying that the $d_b$, $D$ dimensional vectors within $W$ are
linearly independent.
This is analogous to the representation theoretic proof of Schur's lemma.
Furthermore, using the unitarity of $G_g$ and $B_g^b$ we derive from \eqref{eq:GUUB} by multiplying each side of the equation with its conjugate transpose that
\begin{equation}\label{eq:Schur}
 B_g^b\left(W^\dagger W\right)=\left(W^\dag W\right)B_g^b\qquad\forall g\;.
\end{equation}
By Schur's lemma this implies $W^\dag W\propto\mathbbm{1}_{d_b}$ (since $\mathrm{rank}(W^\dagger W)=\mathrm{rank}(W)=d_b\neq0$), confirming 
that indeed all vectors $v^{b}_{q,\beta=1,\dots,d_b}$ are pairwise orthogonal.
This also shows that $v^{b}_{q,\beta}\neq\vec{0}\,\forall\beta$, for all 
non-trivial solutions to \eqref{eq:Mker}.

Finally, we show that all $v^{b}_{q,\beta_1}$'s from within one solution
$w^b_{q}$ are orthogonal to all $v^{b}_{p,\beta_2}$'s from within a distinct solution $w^b_{p}$
for any $\beta_1$, $\beta_2$. Defining the according matrices 
\mbox{$W_{q}:=\mathrm{vec}_D^{-1}(w^b_q)$} and \mbox{$W_{p}:=\mathrm{vec}_D^{-1}(w^b_p)$} 
and following the same steps that led to Eq.~\eqref{eq:Schur} one can show that 
\begin{equation}
 B_g^b\left(W_p^\dagger W_q\right)=\left(W_p^\dag W_q\right)B_g^b\qquad\forall g\;.
\end{equation}
For non-identical solutions $w^b_{p}\,\slashed{\propto}\,w^b_{q}$ the only possible solution to 
this is (again by Schur's lemma) \mbox{$W_p^\dagger W_q=0_{d_b\times d_b}$}.
Altogether this shows the orthogonality of all the $v^{b}_{q,\beta}$'s, and thereby the 
invertibility (and, after appropriate normalization, the unitarity) of $S$.
 \section{Final Remarks}
We have implemented the presented algorithm in a short
\textsc{Mathematica} package for convenience.
The package provides the function \texttt{SBD} that finds a 
unitary solution for $S$ (provided that it exists) given as input two ordered sets of matrices, 
$\{G_1,\dots,G_N\}$ and $\{\mathcal{B}_1,\dots,\mathcal{B}_N\}$,
where the $\mathcal{B}_g$'s are assumed to be block diagonal
with irreducible blocks.

 \section*{Acknowledgements}
I.B. is supported by the IMPRS for Precision Tests of Fundamental Symmetries.

 \appendix
 \section{Example \texorpdfstring{$\mathbf{\D8}$}{D8}}
 \label{sec:D_8}
As an example we discuss the regular representation of the Dihedral group 
\D8.
\renewcommand*{\arraystretch}{1.3}
This finite group has eight elements and is generated by two elements $\mathsf{a}$ and $\mathsf{b}$ 
that fulfill the relations
\begin{align}
    \mathsf{a}^4=\mathsf{e},\quad \mathsf{b}^2=\mathsf{e}, \quad \mathsf{a}\mathsf{b}\mathsf{a}\mathsf{b}=\mathsf{e}\,.
\end{align}
The character table is shown in Tab.~\ref{tab:chartab}.
Generators for the irreducible two-dimensional (2D) representation can be chosen as
\begin{align}\label{eq:rep2}
    \rho_{\textbf{2}}(\mathsf{a}) = \begin{pmatrix}\I &0\\0&-\I \end{pmatrix},\quad\text{and}\quad
    \qquad \rho_{\textbf{2}}(\mathsf{b}) = \begin{pmatrix}0&1\\1&0\end{pmatrix}\;.
\end{align}
The (left-)regular representation acts as
\begin{align}
\mathrm{reg}_\mathsf{a}\;:\qquad\mathsf{g}\mapsto \mathsf{a}\,\mathsf{g}\qquad\forall\mathsf{g}\in\D8\;, \\
\mathrm{reg}_\mathsf{b}\;:\qquad\mathsf{g}\mapsto \mathsf{b}\,\mathsf{g}\qquad\forall\mathsf{g}\in\D8\;. 
\end{align}
These act as permutations on the group elements. 
In a basis $\left\{\mathsf{e,a,a^2,a^3,b,ab,a^2b,a^3b}\right\}$ those
are represented by
\begin{align}\label{eq:reg}
G_\mathsf{a}=\begin{pmatrix}0&1&0&0&0&0&0&0\\ 0&0&1&0&0&0&0&0 \\
0&0&0&1&0&0&0&0 \\ 1&0&0&0&0&0&0&0 \\ 0&0&0&0&0&1&0&0
\\ 0&0&0&0&0&0&1&0 \\ 0&0&0&0&0&0&0&1 \\  0&0&0&0&1&0&0&0\end{pmatrix},\\
G_\mathsf{b} =\begin{pmatrix}0&0&0&0&1&0&0&0\\ 0&0&0&0&0&0&0&1 \\
0&0&0&0&0&0&1&0 \\ 0&0&0&0&0&1&0&0 \\ 1&0&0&0&0&0&0&0
\\ 0&0&0&1&0&0&0&0 \\ 0&0&1&0&0&0&0&0 \\  0&1&0&0&0&0&0&0\end{pmatrix}.
\end{align}
\begin{table}[ht]
\begin{center}
\begin{tabular}{c|ccccc}
\D8& $\{\mathsf{e}\}$ & $\{\mathsf{a}^2\}$ & $\{\mathsf{a},\mathsf{a}^3\}$ & $\{\mathsf{b},\mathsf{a}^2\mathsf{b}\}$ & $\{\mathsf{a}\mathsf{b},\mathsf{a}^3\mathsf{b}\}$ \\
\hline
$\textbf{1}_0$ & $1$ & $1$& $1$& $1$& $1$ \\
$\textbf{1}_1$ & $1$ & $1$& $1$& $-1\hspace{7pt}$& $-1\hspace{7pt}$ \\
$\textbf{1}_2$ & $1$ & $1$& $-1\hspace{7pt}$& $1$ & $-1\hspace{7pt}$\\
$\textbf{1}_3$ & $1$ & $1$& $-1\hspace{7pt}$& $-1\hspace{7pt}$& $1$\\
$\textbf{2}$ & $2$ & $-2\hspace{7pt}$ & $0$ & $0$ & $0$ \\
\hline
\end{tabular}
\end{center}
\caption{Character Table of \D8.}
 \label{tab:chartab}
\end{table}
The regular representation decomposes into irreducible representations as
\begin{equation}
 \mathrm{reg}~=~\mathbf{1}_0 \oplus \mathbf{1}_1 \oplus \mathbf{1}_2 \oplus \mathbf{1}_3 \oplus \mathbf{2}\oplus \mathbf{2}\;.
\end{equation}
Hence, there must be a basis in which $G_\mathsf{a}$ and $G_\mathsf{b}$ are 
block diagonal and given by 
\begin{align}\notag
 B_\mathsf{a} &=\rho_{\textbf{1}_0}(\mathsf{a})\oplus \rho_{\textbf{1}_1}(\mathsf{a})\oplus \rho_{\textbf{1}_2}(\mathsf{a})\oplus \rho_{\textbf{1}_3}(\mathsf{a})\oplus \rho_{\textbf{2}}(\mathsf{a}) \oplus \rho_{\textbf{2}}(\mathsf{a})\;,\\\notag
 B_\mathsf{b} &= \rho_{\textbf{1}_0}(\mathsf{b})\oplus \rho_{\textbf{1}_1}(\mathsf{b})\oplus \rho_{\textbf{1}_2}(\mathsf{b})\oplus \rho_{\textbf{1}_3}(\mathsf{b})\oplus \rho_{\textbf{2}}(\mathsf{b}) \oplus \rho_{\textbf{2}}(\mathsf{b})\;,
\end{align}
with 2D representation matrices given in \eqref{eq:rep2} and 1D 
representations that can be read off from Tab.~\ref{tab:chartab}.

Our algorithm finds a transformation matrix $S$ 
which simultaneously transforms $G_\mathsf{a}$ to $B_\mathsf{a}$,
and $G_\mathsf{b}$ to $B_\mathsf{b}$.
Of course, this is not a real challenge here as for a small number of generators
and small dimensional representations this problem could straightforwardly be solved
by a manual computation. However, our algorithm seamlessly extends to much more 
complicated situations.

The number of generators is $N=2$ and the number of non-identical
blocks is $n=5$ with degeneracies $q_{b=1,\dots,5}=\left\{1,1,1,1,2\right\}$.
For the 1D blocks (commutative part), this reduces to the usual problem 
of finding common eigenvectors, which is simple to solve see e.g.\ \cite{Shemesh:1984}.
In our approach this part of $S$ is determined by finding 
\begin{equation}
\ker(\mathcal{M}^b)=
\ker
\mat{
G_\mathsf{a}-\rho_{\mathbf{1}_{b-1}}(\mathsf{a})\cdot \mathbbm{1}_8\\
G_\mathsf{b}-\rho_{\mathbf{1}_{b-1}}(\mathsf{b})\cdot \mathbbm{1}_8
}\;,
\end{equation}
for $b=1,\dots,4$. These kernels are one-dimensional (as warranted by $q_{b=1,2,3,4}=1$)
and spanned by the orthogonal vectors
\begin{align}
w^{1} &=  \left(1,1,1,1,1,1,1,1\right)^{\mathrm{T}}\;,& \\
w^{2} &=  \left(1,1,1,1,-1,-1,-1,-1\right)^{\mathrm{T}}\;,& \\
w^{3} &=  \left(-1,1,-1,1,-1,1,-1,1\right)^{\mathrm{T}}\;,& \\
w^{4} &=  \left(1,-1,1,-1,-1,1,-1,1\right)^{\mathrm{T}}\;.& 
\end{align}
For the twofold degenerate 2D blocks $(b=5,q_5=2)$ one has to find
\begin{equation}
\ker(\mathcal{M}^5)=
\ker
\mat{
\mathbbm{1}_{2}\otimes G_\mathsf{a} - \rho_\mathbf{2}(\mathsf{a})^{\mathrm{T}} \otimes \mathbbm{1}_{8}\\
\mathbbm{1}_{2}\otimes G_\mathsf{b} - \rho_\mathbf{2}(\mathsf{b})^{\mathrm{T}} \otimes \mathbbm{1}_{8}
}\;.
\end{equation}
In agreement with $q_5=2$, this kernel is two-dimensinal and can be spanned by the two
($D\cdot d_5 = 8\cdot 2$)-dimensional orthogonal vectors 
\begin{align}\notag
w^5_1 &= \left(-\I, 1, \I, -1, 0, 0, 0, 0, 0, 0, 0, 0, -\I, -1, \I, 1\right)^{\mathrm{T}}\;,& \\\notag
w^5_2 &= \left(0, 0, 0, 0, -\I, 1, \I, -1, -\I, -1, \I, 1, 0, 0, 0, 0\right)^{\mathrm{T}}\;.&
\end{align}
According to Proposition $1$, we then find the unitary matrix $S$ 
by joining the vectors $w^b$, applying the inverse vectorization to them, 
and normalizing each column of the resulting matrix.
The result is given by
\begin{equation}
    S=\frac{1}{\sqrt{8}}\left(
\begin{array}{cccccccc}
 1 & 1 & -1 & 1 & -\I\sqrt{2} & 0 & 0 & -\I\sqrt{2} \\
 1 & 1 & 1 & -1 & \sqrt{2} & 0 & 0 & -\sqrt{2} \\
 1 & 1 & -1 & 1 & \I\sqrt{2} & 0 & 0 & \I\sqrt{2} \\
 1 & 1 & 1 & -1 & -\sqrt{2} & 0 & 0 & \sqrt{2} \\
 1 & -1 & -1 & -1 & 0 & -\I\sqrt{2} & -\I\sqrt{2} & 0 \\
 1 & -1 & 1 & 1 & 0 & -\sqrt{2} & \sqrt{2} & 0 \\
 1 & -1 & -1 & -1 & 0 & \I\sqrt{2} & \I\sqrt{2} & 0 \\
 1 & -1 & 1 & 1 & 0 & \sqrt{2} & -\sqrt{2} & 0 \\
\end{array}
\right).
\end{equation}
It is straightforward to check that this matrix is unitary and satisfies \eqref{eq:1} for 
$G_{\mathsf{a,b}}$ and $B_{\mathsf{a,b}}$.
 \section{\boldmath Details on \texorpdfstring{$\mathrm{dim}\,\mathrm{ker}\,\mathcal{M}^b=q_b$}{} \unboldmath}
 \label{sec:dimKernM}
Here we demonstrate that $\dim\ker\mathcal{M}^b=q_b$. 
We first show $\dim\ker\mathcal{M}^b\geq q_b$, and then $\dim\ker\mathcal{M}^b \leq q_b$.

Part 1: $\dim\ker\mathcal{M}^b\geq q_b$. 
Let $V$ denote the space on which the matrices $G_g$ act. 
By assumption, there are $q_b$ copies of the common invariant subspace $V^b$, associated with the blocks $B_g^b$, within $V$. 
That is
\begin{equation}
V\supset V^{b}_1\oplus\dots\oplus V^{b}_{q_b}\;.
\end{equation}%
We now establish that each of these $q_b$ 
invariant subspaces 
defines a non-trivial vector in $\ker\mathcal{M}^b$, and that those $q_b$ 
vectors are pair-wise orthogonal.

Each invariant subspace is spanned by a set of basis vectors 
$\{v^b_{q,\beta=1,\dots ,d_b}\}$.
Then, per definition of an invariant subspace,
\begin{equation}\label{eq:invariantsubspacesIB}
\begin{split}
    G_g\,v^b_{q,1} & =\left[\tilde{B}_g^b\right]_{11}\,v^b_{q,1} + \dots + \left[\tilde{B}_g^b\right]_{d_b1}\,v^b_{q,d_b}\,, \\
    \vdots   & \qquad \vdots  \\
    G_g\,v^b_{q,d_b} & = \left[\tilde{B}_g^b\right]_{1d_b}\,v^b_{q,1} + \dots + \left[\tilde{B}_g^b\right]_{d_bd_b}\,v^b_{q,d_b}\,,
\end{split}
\quad 
\begin{split}
    \forall g\,,
\end{split}\raisetag{20pt}
\end{equation}
where the $[\tilde{B}_g^b]_{\alpha\beta}$ are arbitrary expansion coefficients.
Clearly, we can always choose a basis for $V^b_q$ in which 
$\tilde{B}_g^b=B_g^b$ of Eq.~\eqref{eq:sumofblocks}.
Working in such a basis and arranging $\{v^b_{q,\beta=1,\dots ,d_b}\}$ 
into a $w^q_b$ according to Eq.~\eqref{eq:winv},
one finds that Eq.~\eqref{eq:invariantsubspacesIB} is nothing but the spelled-out version of 
Eq.~\eqref{eq:Mker}. Thus, $w^b_q\in\ker\mathcal{M}^b$ by construction. 
Furthermore, $w^b_{q_1}\perp w^b_{q_2}$ for  $q_1\neq q_2$,
since
\begin{equation}
\braket{w_{q_1}^b,w_{q_2}^b}
= \sum_{\alpha=1}^{d_b} \braket{v^b_{q_1,\alpha},v_{q_2,\alpha}^b}=\sum_{\alpha=1}^{d_b}0=0\,,
\end{equation}
where we have used that $V^b_{q_1}\perp V^b_{q_2\neq q_1}$ by assumption.
Hence, each copy of the invariant subspace provides an independent solution to \eqref{eq:Mker}, 
implying that there are at least $q_b$ orthogonal vectors in $\ker\mathcal{M}^b$.

Part 2: $\dim\ker\mathcal{M}^b \leq q_b$.
Assume $\dim\ker\mathcal{M}^b=Q$. Then we can find $Q$ linearly independent solutions to \eqref{eq:Mker}.
Each of those can be transformed to a \mbox{$(D\times d_b)$}-dimensional left-invertible matrix \mbox{$W^b_q:=\mathrm{vec}_D^{-1}(w^b_q)$}
(see the discussion around Eq.~\eqref{eq:GUUB}). 
Furthermore, since all columns of all $W^b_{q=1,\dots,Q}$ are pair-wise orthogonal these matrices can straightforwardly 
be combined to a $(D\times d_b\cdot Q)$-dimensional left-invertible matrix $W^b_Q$ that fulfills the equation 
\begin{equation}
 {(W^b_{Q})}^{-1}\,G_g\,W^b_Q=\left(B^b_g\right)^{\oplus Q} \qquad\forall g\;.
\end{equation}
Hence, $Q\leq q_b$, as there are, by assumption, exactly $q_b$ copies of $B^b_g$ in $G_g$.
\vfill\null
\bibliography{bibliography}

\end{document}